\documentclass[conference,10pt]{IEEEtran}
\IEEEoverridecommandlockouts
\usepackage{cite}
\usepackage{amsmath,amssymb,amsfonts}
\usepackage{algorithmic}
\usepackage{graphicx}
\usepackage{textcomp}
\usepackage{xcolor}
\usepackage{enumitem}
\newcommand{\norm}[1]{\left\lVert#1\right\rVert}
\def\BibTeX{{\rm B\kern-.05em{\sc i\kern-.025em b}\kern-.08em
    T\kern-.1667em\lower.7ex\hbox{E}\kern-.125emX}}
\begin{document}

\title{Cell-Free Multi-User 
 MIMO Equalization via In-Context Learning\\

\thanks{
	Matteo Zecchin and Osvaldo Simeone are with the King’s Communications, Learning and Information
	 Processing (KCLIP) lab within the Centre for Intelligent Information Processing Systems (CIIPS), Department of Engineering, King’s College London, London WC2R 2LS, U.K. (e-mail: matteo.1.zecchin@kcl.ac.uk;
	 osvaldo.simeone@kcl.ac.uk).
	Kai Yu is with the School of Electronic Science and Engineering, Nanjing
	 University, Nanjing, China, 210023 (e-mail: kaiyu@smail.nju.edu.cn).
	
	The work of M. Zecchin and O. Simeone was supported by the European Union’s Horizon Europe project CENTRIC (101096379). The work of O. Simeone was also supported by the Open Fellowships of the EPSRC (EP/W024101/1) by the EPSRC project  (EP/X011852/1), and by Project REASON, a UK Government funded project under the Future Open Networks Research Challenge (FONRC) sponsored by the Department of Science Innovation and Technology (DSIT).}
}
\author{
    \IEEEauthorblockN{Matteo Zecchin, Kai Yu and Osvaldo Simeone}
}

\author{\IEEEauthorblockN{Matteo Zecchin}
\IEEEauthorblockA{KCLIP lab, CIIPS\\Department of Engineering \\
King’s College London\\
London, UK}
\and
\IEEEauthorblockN{Kai Yu}
\IEEEauthorblockA{School of Electronic Science and Engineering \\
Nanjing University\\
Nanjing, China }
\and
\IEEEauthorblockN{Osvaldo Simeone}
\IEEEauthorblockA{KCLIP lab, CIIPS\\ Department of Engineering \\
	King’s College London\\
	London, UK}
}
\maketitle

\begin{abstract}

Large pre-trained sequence models, such as transformers, excel as few-shot learners capable of in-context learning (ICL). In ICL, a model is trained to adapt its operation to a new task based on limited contextual information, typically in the form of a few training examples for the given task. Previous work has explored the use of ICL for channel equalization in single-user multi-input and multiple-output (MIMO) systems. In this work, we demonstrate that ICL can be also used to tackle the problem of multi-user equalization in cell-free MIMO systems with limited fronthaul capacity.  In this scenario, a task is defined by channel statistics, signal-to-noise ratio, and modulation schemes. The context encompasses the users' pilot sequences, the corresponding quantized received signals, and the current received data signal.  Different prompt design strategies are proposed and evaluated that encompass also large-scale fading and modulation information. Experiments demonstrate that ICL-based equalization provides estimates with lower mean squared error as compared to the linear minimum mean squared error equalizer, especially in the presence of limited fronthaul capacity and pilot contamination.
\end{abstract}


\section{Introduction}

\begin{figure}
    \centering
    \includegraphics[width=0.46\textwidth]{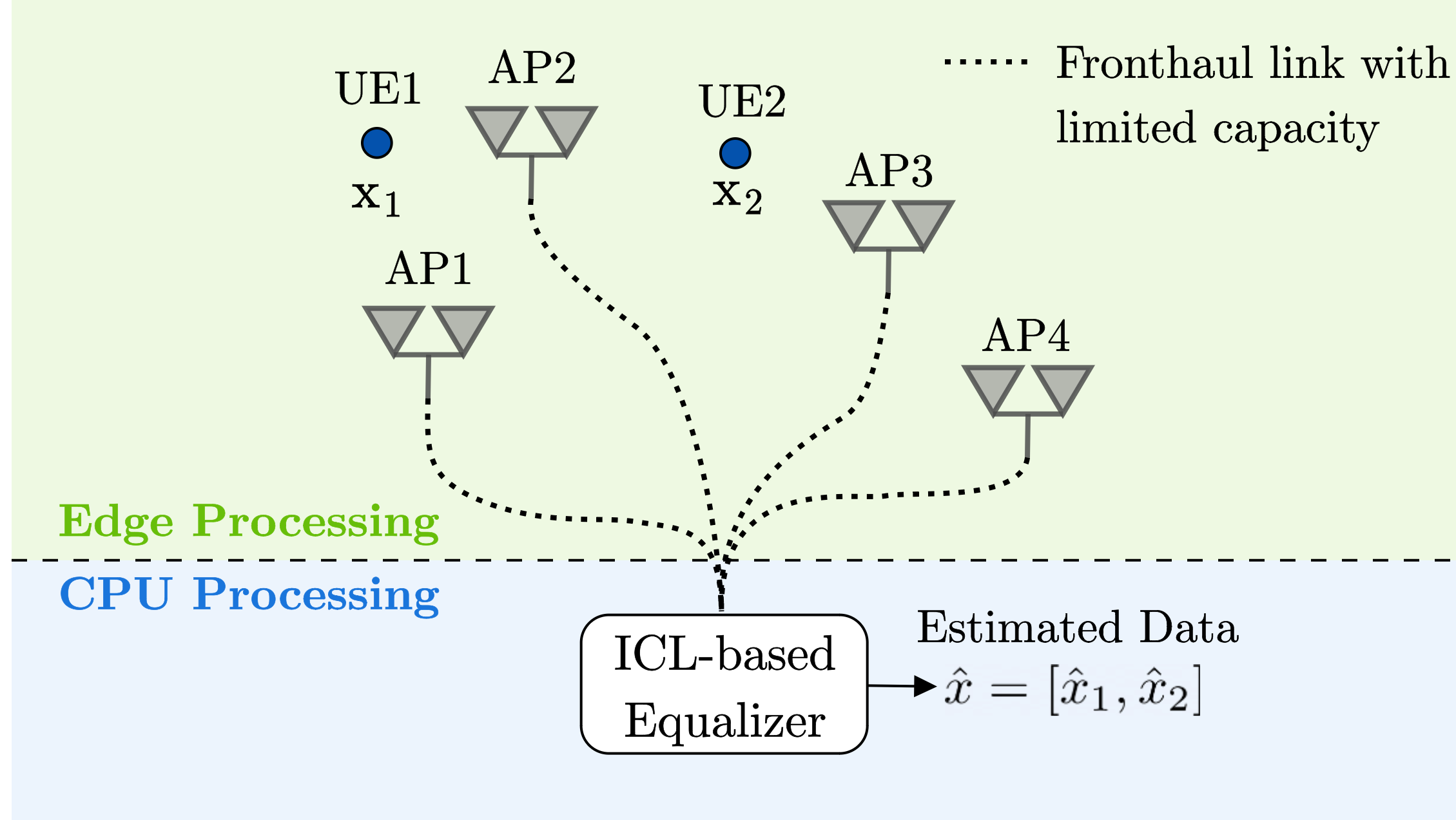}
    \caption{ ICL-based channel equalization for cell-free MIMO systems.}
    \label{fig:Setup}
\end{figure}

Artificial intelligence (AI) models are envisioned to play a central role in next-generation wireless systems, particularly with the advent of disaggregated radio access networks (RANs) \cite{groen2023implementing}. The successful integration of AI models in communication systems hinges on their ability to adapt their operating conditions based on limited contextual information  \cite{simeone2020learning,chen2023learning}. For instance, at the RAN, intelligent wireless networks should be capable of updating their internal operations by using limited pilot sequences to ensure satisfactory performance despite varying operating conditions \cite{raviv2023modular,nikoloska2022modular}.

Recent works have explored different strategies to achieve adaptability. Meta-learning, or learning to learn, has been demonstrated to enable fast adaptation of AI modules based on limited examples \cite{chen2023learning,simeone2022machine}. However, a primary drawback of meta-learning is its requirement for frequent model updates and its sensitivity to hyperparameter choices. More recently, in-context learning (ICL) based on pre-trained sequence models has emerged as a promising alternative. ICL refers to the capacity of sequence models to adapt based on a prompt that includes task information in the form of training samples, without requiring fine-tuning of the model parameters \cite{dong2022survey,Garg2022WhatCT,zhang2023trained}. This learning paradigm has been recently explored for the development of adaptive AI receivers that can adjust based on received pilot sequences \cite{rajagopalan2023transformers, zecchin2023context}.

%performs signal processing and adapts the network operation based on the received pilot sequences from the APs \cite{}. A significant challenge in operating cell-free MIMO systems lies

Cell-free MIMO is a networking paradigm that follows the cloud RAN principle of centralizing signal processing for access points (APs) distributed across a designated deployment area \cite{quek2017cloud,nayebi2015cell,bjornson2019making}. To this end, as shown in Figure 1, a central processing unit (CPU) is connected to the APs via limited-capacity fronthaul links. In the uplink, the need for the APs to quantize the received signals causes the central processor to receive non-linearly distorted signals, posing significant challenges to the design of detection algorithms  \cite{ibrahim2021uplink,masoumi2019performance}.

\subsubsection*{Contributions} This study explores the use of ICL to address the challenge of channel equalization in multi-user cell-free MIMO systems subject to fronthaul capacity constraints.  ICL is leveraged to adapt the operation of a transformer-based equalizer based on limited contextual information without the need for explicit model updates.  As illustrated in Figure 2, the proposed ICL-based maps transmitted and received, quantized, pilot signals, along with the current received data signal, to an estimate of the current data symbol. 
\begin{itemize}[noitemsep, topsep=0pt,wide=0pt]
    \item We introduce  ICL-based equalization for multi-user cell-free MIMO systems under limited fronthaul capacity. Previous work considered only the simpler cellular single-user MIMO systems \cite{zecchin2023context,rajagopalan2023transformers}.
    \item We propose and evaluate different prompt design strategies encompassing large-scale fading and modulation information.
    \item Through empirical results, we evaluate the performance of ICL-based equalization under various fronthaul capacity constraints and levels of pilot contamination. Our results highlight the importance of prompt design and the  superior performance  of ICL as compared to  centralized linear minimum mean squared error (LMMSE) equalization \cite{bashar2020uplink}.
\end{itemize}

\subsubsection*{Notation}  Given a matrix $X$ we use $\left[\mathnormal X \right]_{i,j}$ to denote the entry $(i,j)$, while we refer to the $i$-th column by $\left[\mathnormal X \right]_{i}$. For a vector $x$,  $[\mathnormal x]_i$ denotes the $i$-th entry. Superscripts $^{\textrm{T}}$, $^{\dag}$ and $^{\textrm{H}}$ refer respectively to the transpose, conjugate, and conjugate transpose operations.
We use $\mathcal{CN}({\mu},\mathnormal R) $ to refer to  complex  circularly-symmetric Gaussian distribution with mean ${\mu}$ and covariance $\mathnormal R$.

\section{System Model}

\label{sec:system_model}
 As illustrated in Figure \ref{fig:Setup}, we investigate the problem of channel equalization in a cell-free multi-user MIMO system with limited fronthaul capacity. The network consists of $K$ single-antenna user equipments (UEs) and of $L$ access points (APs), each equipped with $N$ antennas, which are coordinated by a central processing unit (CPU). Following 5G and Open Radio Access Networks (O-RAN) terminologies, APs may be also referred to as Radio Units (RUs), while the CPU corresponds to a Distributed Unit (DU) \cite{demir2024cell}. 

The wireless channel between UE $k$ and AP $l$ is denoted by $\mathnormal{h}_{l,k}\in\mathbb{C}^{N}$. We assume a Rayleigh block-fading model, wherein the channel $\mathnormal{h}_{l,k}$ remains constant within each coherence block of length $T$ and follows a correlated Rayleigh distribution; i.e.,
\begin{align}
    \mathnormal{h}_{l,k}\sim \mathcal{CN}(0,\mathnormal{R}_{l,k}),
    \label{eq:rayleigh_ch}
\end{align}
where $\mathnormal{R}_{l,k}$ is the $N\times N$ spatial correlation matrix. The large-scale fading coefficient of the channel $\mathnormal{h}_{l,k}$ is denoted by $r_{l,k}=\textrm{tr}(\mathnormal{R}_{l,k})/N$; and the large-scale fading coefficients for the channels  from UE $k$ to all APs are collected in vector  $\mathnormal{r}_{k}=[r_{1,k},\dots,r_{L,k}]$.

We consider pilot-aided channel equalization, in which,  during the initial $T_p < T$ channel uses, UEs transmit pilot sequences, followed by data transmission for the remaining $T_d = T - T_p$ channel uses. In the pilot transmission phase, each user $k$ is assigned a pilot sequence ${\phi}_{p_k}$ of length $T_p$ symbols and normalized energy  $\norm{\phi_i}^2 = T_p$. The pilot sequence ${\phi}_{p_k}$ is  chosen from a set $\{\phi_1, \dots, \phi_{T_p}\}$ of $T_p$ orthogonal pilot sequences. The  transmitted pilot sequences assigned to all $K$ UEs are collected into matrix $ {\Phi}=\left[\phi_{p_1},\dots,\phi_{p_K}\right]^{\mathrm{T}}\in \mathbb{C}^{K\times T_p}$.

\begin{figure}
    \centering
    \includegraphics[width=0.46\textwidth]{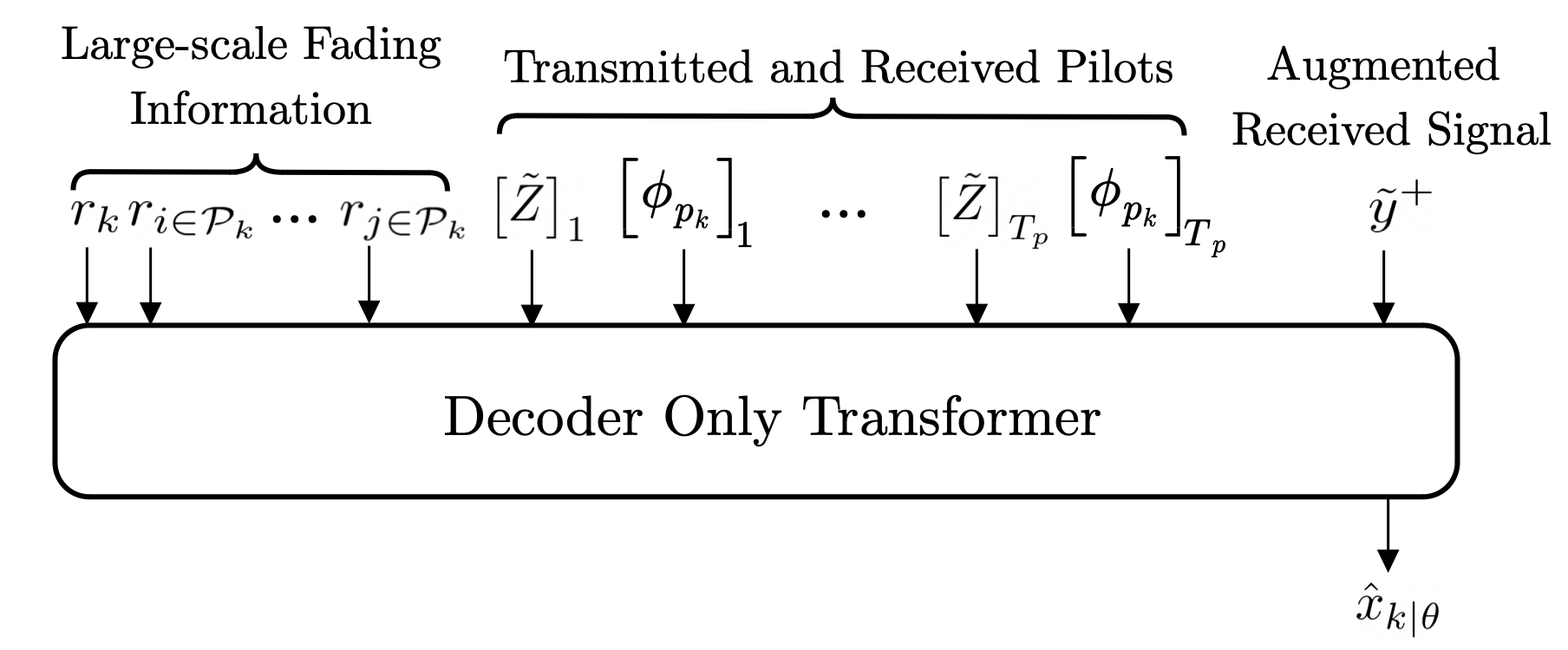}
    \caption{Illustration of the ICL-equalizer based on decoder-only transformers. }
    \label{fig:icl-equalizer}
\end{figure}

As in standard cell-free massive MIMO models \cite{nayebi2015cell},  pilot reuse is allowed, as more than one user can be assigned to the same pilot sequence. The set of UEs assigned to the same pilot as user $k$, excluding itself, is denoted by $\mathcal{P}_k\subseteq \{1,\dots,K\}$. As a result of the simultaneous pilot transmission by all UEs, the received signal at  each AP $l$ across the $T_p$ channel uses is given by 
\begin{align}
    \mathnormal{Z}^{}_l=\sum^K_{k=1}\mathnormal{h}_{l,k}\phi^{\mathrm{T}}_{p_k}+\mathnormal{N}^{}_l,
    \label{eq:received_pilots}
\end{align}
where $\mathnormal{N}^{}_l\in \mathbb{C}^{N\times T_p}$ is the receiver noise matrix with independent complex Gaussian entries $\left[\mathnormal{N}^{}_l\right]_{i,j}\sim \mathcal{CN}(0,\sigma^2)$.

During each channel use of the data transmission phase, each UE $k$ transmits a data symbol $x_k\in\mathcal{X}_k$ chosen uniformly at random from a constellation $\mathcal{X}_k$. The average transmit power is normalized as  $\mathbb{E}[\norm{x_k}^2]=1$; and we denote the concatenation of the transmitted data symbols as 
$\mathnormal{x}=\left[x_1,\dots,x_K\right]^{\mathrm{T}}\in \mathbb{C}^{K}$.  
Accordingly, for each channel use of the data phase,  the received  signal at AP $l$ is
\begin{align}
     \mathnormal{y}_l=\sum^K_{k=1}\mathnormal{h}_{l,k}x_k+\mathnormal{n}_l
    \label{eq:received_data}
\end{align}
where $\mathnormal{n}_l\sim  \mathcal{CN}(0,\sigma^2\mathnormal{I}_N)$ is the receiver noise vector.

We consider centralized equalization at the CPU, where each AP acts as a relay by forwarding the received pilots $\mathnormal{Z}^{}_l$ and data $\mathnormal{y}_l$ to the CPU for processing. The fronthaul links connecting APs to the CPU have a limited capacity of $b$ bits per symbol. To adhere to the fronthaul communication constraints, each AP $l$ applies a $b$-bit quantizer $\mathcal{Q}_b(\cdot)$ entry-wise, and separately to real and imaginary components, to the signals $\mathnormal{Z}^{}_l$ and $\mathnormal{y}_l$. We denote the quantized received pilots and data symbols as $\tilde{\mathnormal{Z}}^{}_l$ and $\tilde{\mathnormal{y}}_l$ respectively,  the concatenation of all quantized received pilots as $\tilde{\mathnormal{Z}}^{}=[\tilde{\mathnormal{Z}}^{}_1,\dots,\tilde{\mathnormal{Z}}^{}_L]^{\mathrm{T}}\in \mathbb{C}^{N L\times T_p}$, and the concatenation of all quantized received data as $\tilde{\mathnormal{y}}=\left[\tilde{\mathnormal{y}}_1,\dots,\tilde{\mathnormal{y}}_L\right]^{\mathrm{T}}\in \mathbb{C}^{NL\times 1}$.

Overall,  a channel equalization task $\tau$ is specified by the tuple $\tau=\left(\sigma^2, K, \{\mathcal{X}_k\}^K_{k=1},\{\mathnormal{R}_{1,k}\}^K_{k=1},\dots, \{\mathnormal{R}_{L,k}\}^K_{k=1}\right)$, which determines the noise variance $\sigma^2$, the number $K$ of active UEs, the constellations $\{\mathcal{X}_k\}^K_{k=1}$,  and the spatial correlation matrices $\{\mathnormal{R}_{1,k}\}^K_{k=1},\dots, \{\mathnormal{R}_{L,k}\}^K_{k=1}$ of the channels.

\section{Centralized Equalization}
\label{sec:equalization}
In this work, we consider centralized equalization at the CPU, in which the goal is to estimate the data $x$ from the quantized received pilots $\tilde{\mathnormal{Z}}^{}$ and corresponding received signal $\tilde{\mathnormal{y}}$, as well as from task information obtained as a function $f(\tau)$ of the task identifier $\tau$. In this section, we first review the centralized linear minimum mean squared error equalizer (LMMSE) \cite{bjornson2019making,bashar2020uplink}, which serves as a baseline, and then illustrate the proposed ICL-based equalization scheme.

\subsection{Centralized Linear MMSE Equalization }
The centralized LMMSE equalizer performs linear processing of the received data signal $\tilde{\mathnormal{y}}$ based on the received pilots $\tilde{\mathnormal{Z}}^{}$ and using task knowledge $f(\tau)=(\sigma^2,K,\{\mathnormal{R}_{1,k}\}^K_{k=1},\dots, \{\mathnormal{R}_{L,k}\}^K_{k=1})$. The LMMSE equalizer is thus not tailored to specific constellations $\{\mathcal{X}_k\}^K_{k=1}$. 
 Linear processing is optimized  by modeling quantization operation via the Bussgang decomposition, treating channel estimates as exact, and assuming Gaussian data symbols \cite{bjornson2019making}.  Owing to these simplifying assumptions, the resulting centralized LMMSE equalizer is generally suboptimal. 

 Referring to \cite{bashar2020uplink} for details, based on the known channel correlations, the centralized LMMSE equalizer first estimates the channel between each AP $l$ and UE $k$ as $\hat{h}_{l,k}$. Then, the channel estimates are used to derive the optimal linear equalizers under the mentioned assumptions, obtaining the equalized data symbol for each user $k$ as
\begin{align}
    \label{eq:LMMSEeq}
    \hat{x}^{\mathrm{LMMSE}}_k=v^{\mathrm{H}}_{k}\tilde{y},
\end{align} where $v_k$ is the optimized linear equalizer.

\subsection{Centralized ICL-Based Equalization: Architecture} 
\label{sec:ICL_eq}

We now introduce the proposed ICL-based equalizer. As illustrated in Figure 2, ICL equalization is implemented by mapping the transmitted and received pilots, along with the current received data signal and with task information, to the estimated transmitted data by using a decoder-only transformer architecture. Accordingly, the estimate of the data point $x_k$ corresponding to the current $NL\times 1$ received signal $\tilde{y}$ for each UE $k$ is obtained via a mapping 
\begin{align}
\hat{\mathnormal{x}}_{k|\theta}=\hat{\mathnormal{x}}_\theta\left(f(\tau),\mathcal{D}_{\tau,k},\tilde{\mathnormal{y}}\right),
\label{eq:icl_estimate}
\end{align}
where $\theta$ is the transformer's parameter vector, while task information $f(\tau)$ and context $\mathcal{D}_{\tau,k}$ are detailed next.

The context $\mathcal{D}_{\tau,k}$ encompasses the UE $k$'s transmitted pilot sequence and the corresponding sequence of received signals over  $T_p$ channel uses as
\begin{align}
\mathcal{D}_{\tau,k}=\{([\tilde{\mathnormal{Z}}^{}]_i,[{\phi_{p_k}}]_i)\}^{T_p}_{i=1},
\label{eq:example_sequence}
\end{align} where the $NL\times 1$ vector  $[\tilde{\mathnormal{Z}}^{}]_i$  and the scalar $[\phi_{p_k}]_i$ correspond to received and transmitted pilots at the $i$-th channel use.  In addition to the received pilots $\tilde{\mathnormal{Z}}^{}$, the ICL-equalizer is provided with task information $f(\tau)=\left(\{\mathcal{X}_k\}^K_{k=1},\{r_k\}^K_{k=1}\right)$, which, unlike the centralized LMMSE estimator, includes only information about the large-scale fading coefficients and the modulation scheme adopted by UEs.
Task information $f(\tau)$ is embedded into the sequence (\ref{eq:example_sequence}) as explained in Section \ref{sec:prompt_design}.

\subsection{Centralized ICL-Based Equalization: Prompt Design}

\label{sec:prompt_design}
 As discussed, the estimate $ \hat{\mathnormal{x}}_{k|\theta} $ in (\ref{eq:icl_estimate}) depends on  task information $ f(\tau) $, context $ \mathcal{D}_{\tau,k} $, and  received signal $\tilde{y}$. We now discuss how to specify the prompt $(f(\tau),\mathcal{D}_{\tau,k},\tilde{\mathnormal{y}})$ in terms of a sequence of tokens.

 First, as shown in Figure 1, the prompt contains large-scale fading information, Specifically, the prompt includes the large-scale fading coefficient  $\mathnormal{r}_k$ for UE $k$ as a token, followed by the large-scale coefficients  $\{\mathnormal{r}_i\}_{i\in \mathcal{P}_k}$  of the UEs associated with the same pilot sequence as UE $k$ as separate additional tokens. These additional tokens provide statistical information about the channels associated with UEs with colliding pilots, which  can be leveraged by the transformer  to mitigate the effect of pilot contamination.

 The large-scale fading tokens are followed by the $2T_p$  tokens corresponding to the context  $ \mathcal{D}_{\tau,k}$. These are obtained as in (\ref{eq:example_sequence}) by two tokens per channel use. Finally, the received signal $\tilde{y}$ is augmented with information about the constellation $\mathcal{X}_k$.  To this end, assuming that there are $M$ possible constellations, the constellation $\mathcal{X}_k$ is described by an index $m(\mathcal{X}_k)\in \{0,\dots,M-1\}$, which is concatenated with the received signal $\tilde{y}$ to define the token $\tilde{\mathnormal{y}}^+=[\tilde{\mathnormal{y}},m(\mathcal{X}_k)]$. 

 Tokens that include complex signals are mapped to real vectors by concatenating real and imaginary components. Furthermore, since the prompt, as described in this subsection, includes vectors with different dimensions, zero padding is used to ensure that all tokens have the dimension corresponding to the largest vector.

\subsection{Pre-Training} 

The parameter vector $\theta$ of the ICL-based equalizer is optimized offline based on a pre-training set $\mathcal{T}_{\mathrm{tr}}$ including data for  $N_{\mathrm{tr}}$ tasks sampled i.i.d. from a common distribution. Specifically, for each task $\tau\in\mathcal{T}_{\mathrm{tr}}$,  we assume access to the task information $f(\tau)$, as well as to a number $N_{\mathrm{ex}}$ of i.i.d.  generated examples including   contexts $\mathcal{D}_{\tau,k}$ for all users $k\in \{1,...,K\}$ and a test pair $(\mathnormal{x},\tilde{\mathnormal{y}})$. Using these data, during the pre-training phase, the ICL-based equalizer parameter vector $\theta$ is optimized to minimize the empirical mean squared error averaged over the pre-training tasks $\mathcal{T}_{\mathrm{tr}}$; i.e., the training loss is
\begin{align}
    \mathcal{L}(\theta)&=\sum_{\tau\in \mathcal{T}_{\mathrm{tr}}} \sum_{k=1}^K \mathbb{E}\big[\norm{\hat{\mathnormal{x}}_{k|\theta}-\mathnormal{x}_k}^2\big],
\end{align} where the expectation is evaluated using the $N_{\mathrm{ex}}$ examples available for each task $\tau$. 
Note that the target $\mathnormal{x}$ is available only during training, and it is not known during inference.

\section{Experiments}
This section considers a cell-free MIMO system with varying fronthaul capacity constraints and different pilot contamination scenarios with the aim of comparing the performance of ICL-equalization to the centralized LMMSE benchmark \cite{bjornson2019making}. We also include the performance of a standard meta-learning algorithm, MAML, as an additional baseline \cite{finn2017model,chen2023learning}, as well as the performance of the LMMSE equalizer with an infinite fronthaul capacity \cite{bjornson2019making}.  The simulation code is available at the link \textit{https://github.com/kclip/Cell-free-MIMO-ICL.git}.

\subsection{Simulation Set-Up}
\label{sec:sim_setup}
We adopt a simulation set-up similar to \cite{bjornson2019making}, and consider a square deployment of size $1\times 1$ km with $L=4$ APs with $N=2$ antennas each. For each task $\tau$, a random number of UEs $K\sim \mathcal{U}[1,4]$ are placed in the deployment area by sampling the spatial coordinates uniformly at random. Communication is at the 2 GHz carrier frequency, and path loss and correlated fading are modeled based on the 3GPP Urban Microcell standard \cite{3gpp}. The spatial correlation is generated using the Gaussian local scattering model \cite{bjornson2017massive}, and for training we assume a noise power $\sigma^2=-146$ dB, which corresponds to an average signal-to-noise ratio (SNR) $\mathbb{E}[\norm{h_{l,k}}^2]/(N\sigma^2)$ of 24 dB.

Pilot sequences of length $T_p=8$ are generated from a Walsh-Hadamard matrix. During data transmission, each UE generates symbols by choosing uniformly at a random a constellation $\mathcal{X}_k$ from a set of possible constellations that includes BPSK, 8-PSK, 4-QAM, 16-QAM, and 64-QAM modulations.
To satisfy the limited fronthaul capacity, each AP quantizes the received signal using a $b$-bit non-uniform quantizer,  which is designed by minimizing the average mean squared quantization error when  assuming that the components of the received signal vector are Gaussian with zero mean and power evaluated  using the channel statistics.

The number of tasks used to train the ICL-based equalizer is $N_{\mathrm{tr}}=8192$ and the number of examples per each task is $N_{\mathrm{ex}}=1024$. All training examples are generated by assuming that pilots are selected uniformly without replacement for all UEs.

The ICL-equalizer is implemented using a 4-layer transformer model with absolute positional encoding, an embedding dimension of 64, and 4 attention heads. For the MAML-equalizer, we employ a 3-layer multilayer perceptron with 50 hidden units per layer.

\begin{figure}
    \centering
    \includegraphics[width=0.48\textwidth]{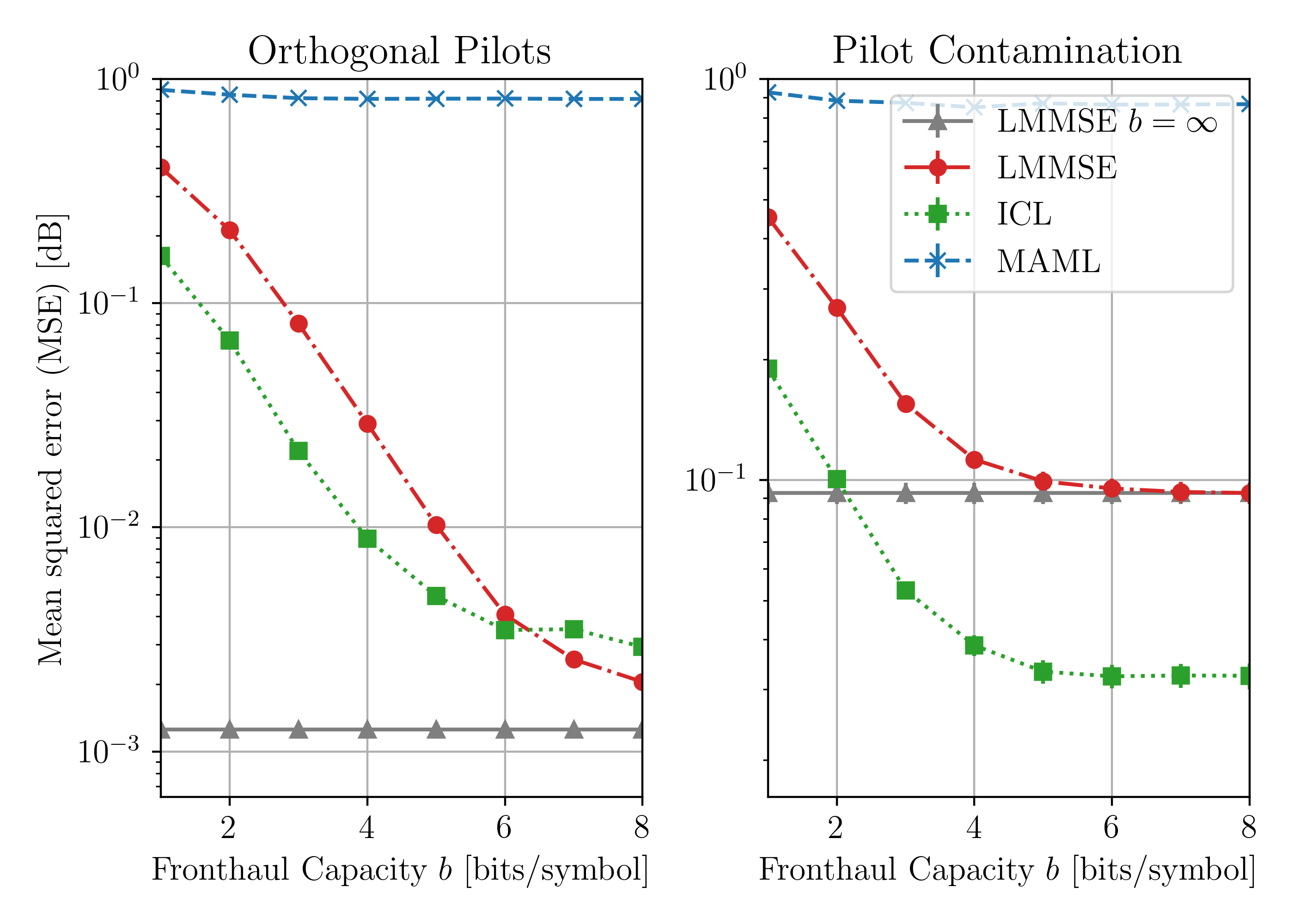}
    \caption{Test mean squared error as a function of the fronthaul capacity $b$ for an orthogonal pilot assignment on the left panel, and for a randomized assignment with pilot reuse on the right ($\text{SNR}=24$ dB). }
    \label{fig:BitsvsMSE}
\end{figure}

\subsection{Results}

In Figure \ref{fig:BitsvsMSE}, we investigate the effect of the limited fronthaul capacity.  For testing, we consider two scenarios: one in which orthogonal pilots are assigned to the UEs (left panel), and one in which pilots are randomly assigned and reused as assumed in the training data set  (right panel). 

With no pilot contamination, as shown in the left panel in Figure \ref{fig:BitsvsMSE}, for small fronthaul capacity $b$, the ICL-based equalizer offers superior MSE performance as compared to the centralized LMMSE equalizer. As the capacity $b$ increases, the performance of both  ICL-based equalizer and  LMMSE equalizer improves, approaching that of the LMMSE equalizer with infinite fronthaul capacity. In contrast,  MAML-based equalization fails to adapt given the available pilot information. This failure may be ascribed to the difference between context data used for adaptation, which is based on pilot transmission, and testing data, which amounts to transmissions from given constellations. In fact, MAML does not have a mechanism to incorporate task descriptors in the adaptation process. 

The right panel of Figure \ref{fig:BitsvsMSE} shows that pilot contamination can have a significant impact on the performance of both equalizers. Importantly, while the centralized LMMSE performance approaches the performance of centralized LMMSE with infinite fronthaul capacity, the ICL-based equalizer converges to a lower MSE value, outperforming both baselines. This showcases the capacity of ICL to adapt the operation of the equalizer based on limited pilot information, while the performance of LMMSE is limited by the simplifying design assumption made in deriving the linear equalizer (\ref{eq:LMMSEeq}).

\begin{figure*}
    \centering
    \includegraphics[width=0.92\textwidth]{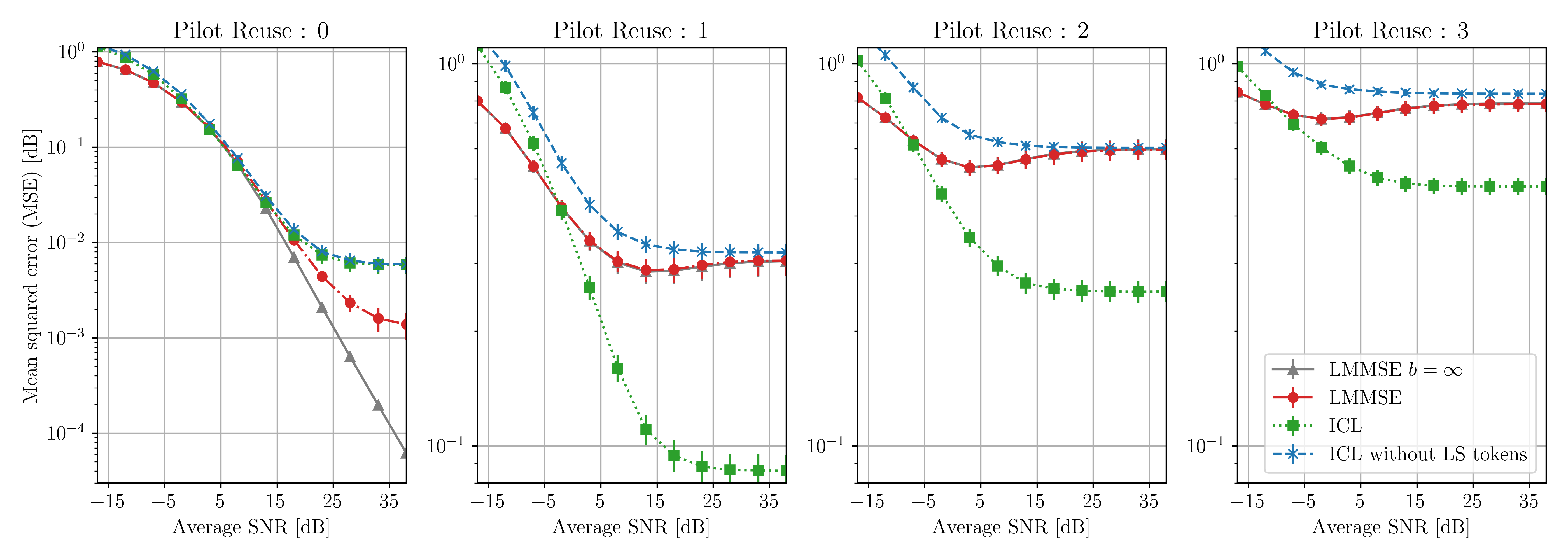}
    \caption{Test mean squared error as a function of the average SNR and for different levels of pilot reuse (Fronthaul capacity $b=8$ bits per symbol).}
       \label{fig:pilot_reuse}
\end{figure*}

To further study the effect of pilot contamination, we now consider a testing scenario with increasing pilot reuse. Specifically, Figure \ref{fig:pilot_reuse} examines four different settings in which the same pilot sequence is reused 0, 1, 2, or 3 times. Assuming $K=4$ UEs, the extreme cases of 0 and 3 correspond to the allocation of orthogonal pilots and to the reuse of the same pilot sequence for all users, respectively. We set $b=8$ bits per symbol, and we vary the average SNR by changing the noise power $\sigma^2$. The figure confirms that,  for large SNR values, there exists a significant gap between the performance of ICL and LMMSE equalization, even when the latter is evaluated with an infinite fronthaul capacity.

 To highlight the importance of prompt design in obtaining the discussed performance levels, we also evaluate the performance of an ICL-based equalizer that is prompted without the large-scale fading tokens in Figure 2. Figure \ref{fig:pilot_reuse} shows that, without pilot reuse, the performance of ICL-based equalization with and without large-scale tokens is comparable.   However, with increasing pilot contamination, the importance of large-scale fading information becomes evident, ensuring the mentioned advantages of ICL-based equalization over existing baselines.

\section{Conclusion}
In this work, we have presented a transformer-based equalizer for multi-user cell-free MIMO systems with limited fronthaul capacity. The proposed equalizer estimates uplink data based on pilot sequences, large-scale fading coefficients, and modulation information used as parts of a prompt to a decoder-only transformer. The transformer implements in-context learning (ICL) to adapt its operation without requiring any fine-tuning. As compared to a centralized  LMMSE equalizer and to conventional meta-learning, ICL-based equalization offers lower-MSE estimates, especially in the most challenging settings with limited fronthaul capacity and with pilot contamination.  Directions for future work may include understanding the generalization performance of ICL and exploring other applications of ICL to wireless communication systems \cite{maatouk2023large,wang2022transformer}.

\bibliographystyle{IEEEtran}
\bibliography{biblio}
\end{document}